\documentclass[preprint,12pt, a4paper]{elsarticle}

\usepackage{amsmath,amsbsy}
\usepackage{longtable}

\usepackage{subfig}
\usepackage{hyperref} 
\biboptions{round, authoryear}

\usepackage{tikz}

\usepackage[nomarkers, nolists,figuresonly]{endfloat}

\journal{Astrophysical Journal}

\begin{document}

	\begin{frontmatter}

		\title{Determining fireball fates using the $\alpha$--$\beta$ criterion}
		
		\author{Eleanor K. Sansom\corref{cor1} }
		\author{Maria Gritsevich\corref{cor2} }
		\author{Hadrien A. R. Devillepoix\corref{cor1}}
		\author{Trent Jansen-Sturgeon\corref{cor1} }
		\author{Patrick Shober \corref{cor1}}
		\author{Phil A. Bland\corref{cor1}}
		\author{Martin C. Towner\corref{cor1}}
        \author{Martin Cup\'ak\corref{cor1}}
        \author{Robert M. Howie\corref{cor1}}
        \author{Benjamin A. D. Hartig\corref{cor1}}

		\cortext[cor1]{Space Science and Technology Centre, Curtin University, GPO Box U1987, Perth, WA 6845, Australia}
		\cortext[cor2]{Department of Physics, Helsinki University, Finland\\ Finnish Geospatial Research Institute (FGI), Masala, Finland\\ Institute of Physics and Technology, Ural Federal University, Ekaterinburg, Russia}
		
		
		\begin{abstract}

As fireball networks grow, the number of events observed becomes unfeasible to manage by manual efforts. Reducing and analysing big data requires automated data pipelines.
Triangulation of a fireball trajectory can swiftly provide information on positions and, with timing information, velocities. However, extending this pipeline to determine the terminal mass estimate of a meteoroid is a complex next step. Established methods typically require assumptions to be made of the physical meteoroid characteristics (such as shape and bulk density). To determine which meteoroids may have survived entry there are empirical criteria that use a fireball's final height and velocity -- low and slow final parameters are likely the best candidates. We review the more elegant approach of the dimensionless coefficient method. Two parameters, $\alpha$ (ballistic coefficient) and $\beta$ (mass-loss), can be calculated for any event with some degree of deceleration, given only velocity and height information. $\alpha$ and $\beta$ can be used to analytically describe a trajectory with the advantage that they are not mere fitting coefficients; they also represent the physical meteoroid properties. This approach can be applied to any fireball network as an initial identification of key events and determine on which to concentrate resources for more in depth analyses. We used a set of 278 events observed by the Desert Fireball Network to show how visualisation in an $\alpha$ -- $\beta$ diagram can quickly identify which fireballs are likely meteorite candidates. 
		\end{abstract}
		
			
		
	\end{frontmatter}


\section{Introduction} \label{sec:intro}
Meteorites are examples of planetesimal building blocks and hold invaluable information on early solar system processes. Less than 0.1\% have known pre-impact origins. 
When extraterrestrial material encounters the Earth's atmosphere, a bright phenomenon can be observed as the meteoroid ablates and ionises the atmosphere. If observed from different locations with high precision, these phenomena can be triangulated and their trajectories determined. Dedicated observation networks, such as the Desert Fireball Network in Australia, record the timing along the luminous trajectory to acquire velocity information \citep{howie2017build}. 

The goal of such networks is to determine heliocentric orbits for these bodies as well as establishing if any mass survived atmospheric ablation to impact the Earth's surface. 
Recovering a fresh meteorite minimises terrestrial contamination, and the ability to associate an orbit with this material is of exceptional value. Despite the knowledge obtainable from meteorite samples on Solar System formation and evolution, very few have orbits to provide location context information ($<0.1\%$). Fireball networks are bridging the gap between asteroidal observations and meteoritic analyses by providing this context.

\citet{Whipple1938} details the first multi-station photographic meteor program  from the mid 1930s, designed to determine trajectories and velocities of meteors. Larger fireball networks have been observing the skies since the 1960s \citep{ceplecha1997prairie} and have accumulated large datasets, though those deemed ``unspectacular" were classed as low priority for data reduction \citep{Halliday1996}. There were not enough resources to measure and reduce all observed meteors, and it was an identified bias in flux surveys. Interesting events were assessed to determine if they were candidates for meteorite searches \citep{Halliday1996}. 
Common practice for identifying which meteoroids may have survived entry is by assessing a fireball's final height and velocity -- low and slow final parameters are likely the best candidates. \citet{brown2013meteorites} discuss how this was empirically determined by early studies of meteorite producing fireballs of the MORP \citep{Halliday1989} and the PN \citep{McCrosky1971}. The set of empirically determined conditions for a fireball to produce a meteorite is an end height below $35\,km$ and a terminal velocity below $10\,km\,s^{-1}$ \citep{Halliday1989, brown2013meteorites, Wetherill1981}. This has been used to direct resource focus to the most likely meteorite dropping events.

\subsection{Established methods of identifying meteorite-dropping events}
Despite advances, reducing fireball data to determine terminal mass estimates is still a non-trivial task. Established methods, such as \citet{Sansom2016, Sansom2017, egal2017challenge, Ceplecha2005}, are based on a set of single body aerodynamic equations that require assumptions to be made about the physical properties of the meteoroid, or in some way statistically estimate their values. These unobservable values, such as shape, density and even ablation efficiencies, introduce many degrees of freedom to modelling scenarios.  
More complex Monte Carlo and particle filter techniques can intelligently assess the parameter space to give statistical likelihood of parameter sets (i.e. \citealp{Sansom2017}). However, these methods still require a multivariate solution and require supercomputing resources to run. 

One concise way of assessing the trajectory without assuming any parameters is the dimensionless coefficient method first described by \citet{gritsevich2007approximation}. The method is based on dimensionless equations describing the trajectory introduced by \citet{Stulov1995}. \citet{gritsevich2006} describe the simplified (asymptotic) solution of the method, and the latest, more advanced realisation of the algorithm (including the incorporation of an arbitrary atmospheric model) is well outlined in \citet{Lyytinen2016}.
The ballistic coefficient $\alpha$, and mass loss parameter $\beta$ can be calculated for any event with some degree of deceleration, given only velocity and height information. For meteors showing no deceleration these parameters may be linked to the terminal height of luminous flight \citep{moreno2015new}. These two parameters can be used to analytically describe a trajectory, given an entry velocity ($V_0$). This is similar to the mathematical curve fitting performed by \citet{Jacchia1956}, subsequently improved by \citet{egal2017challenge}, with the added advantage that there is a link to the physical meteoroid parameters through using $\alpha$ and $\beta$ rather than mere fitting coefficients. This link allows more robust conclusions to be made on the incoming body by assessing the groupings of specific $\alpha$--$\beta$ values.
This is also a fast and easy method to implement and run on a large dataset, such has been done by \citet{Gritsevich2009parameters} for both the Prairie Network (PN) and Meteor Observation and Recovery Project (MORP) data. It has also been applied to well-documented meteorite falls including P\u{r}\'{i}bram, Lost City, Innisfree, Neuschwanstein \citep{Gritsevich2008_4falls}, Bunburra
Rockhole \citep{Sansom2015}, Annama \citep{Lyytinen2016}, Park Forest \citep{meier2017park}, and Ko\u{s}ice \citep{gritsevich2017constraining}.

\subsection{Applying the $\alpha$--$\beta$ criterion to DFN events}
Here we calculate the $\alpha$ and $\beta$ parameters for 278 fireballs observed by the Desert Fireball Network (Section \ref{sec:methods}). This is a subset of some 1300+ fireball trajectories triangulated by the DFN, where noticeable deceleration has occurred ($V_f/V_0<80\%$). We then plot these data in a similar fashion to PN and MORP data in \citet{Gritsevich2012}\footnote{Note that fireballs from the PN and MORP surveys were not subject to any deceleration threasholding.} 
The location of events on this plot instantly allows us to identify key events, such as those likely to drop meteorites. This is an under-utilised tool by fireball networks with large datasets to determine such events to concentrate resources for data reduction. Often identifying good meteorite dropping candidates is done by assessing how low and slow a fireball was observed in our atmosphere using the empirical criteria (end height $<35\,km$ and a final velocity $<10\,km\,s^{-1}$ \citealp{Halliday1989, brown2013meteorites, Wetherill1981}). However, such a classification scheme is highly dependent on the equipment used to record a fireball, and the range at which it was observed.
This is also not a rigorous assessment of the event where slope, mass and shape dependencies all come into play 
The $\alpha$--$\beta$ approach may seem over simplified, but led to the fast recovery of both the Annama meteorite \citep{gritsevich2014first, Trigo2015, dmitriev2015orbit, kohout2017annama} and Ozerki\footnote{https://www.lpi.usra.edu/meteor/metbull.php?code=67709} meteorite.

With the statistically large dataset of the DFN, along with PN and MORP data, we aim to establish an $\alpha$--$\beta$ criterion for classifying the possible outcomes of meteoroid atmospheric entry (Section \ref{sec:fall_region}). We are ultimately looking to establish crude criteria for whether further analyses and meteorite searches are worth prioritising.

\section{The $\alpha$--$\beta$ Diagram -- Desert Fireball Network Data}\label{sec:methods}
\begin{figure}[h!]
    \centering
    \includegraphics[width=0.8\textwidth]{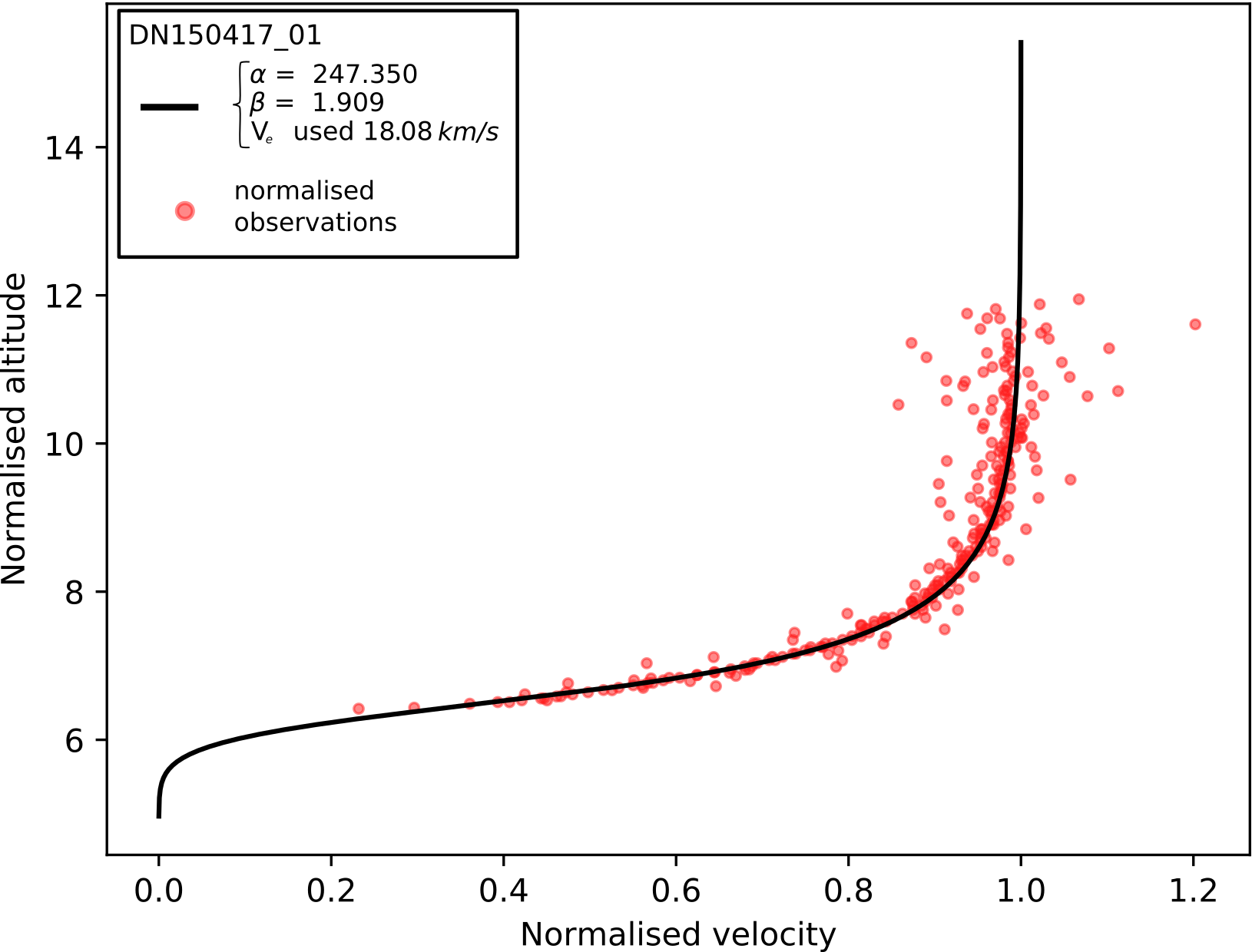}
    \caption{Plot of observational data with velocity normalised to entry velocity $V_0$ and height normalised to the atmospheric scale height ($h_0=7160$\,m). The fit is good despite significant scatter in the data. }
    \label{fig:ab_0xamp}
\end{figure}

Values of $\alpha$ and $\beta$ are calculated using a least squares minimisation of the analytical function (see section 3 of \citealp{Lyytinen2016}, after \citealp{Gritsevich2007morp})  
\begin{equation} 
	y =\ln\alpha+\beta-\ln\cfrac{\Delta}{2} \label{eqn:y}         
\end{equation}
where
$y$ is the height of the meteoroid normalised to the atmospheric scale height ($h_0=7160$\,m), $\Delta$ is a function of the exponential integral ($\Bar{Ei}$) as follows:
\begin{equation*}
\Delta = \Bar{Ei}(\beta)-\Bar{Ei}(\beta v^2),
\end{equation*}
and $v$ the meteoroid velocity normalised by $V_0$.
An example of the fit of this function to observational data is shown in Figure \ref{fig:ab_0xamp}. The code used to generate such figures, and determine $\alpha$ and $\beta$ values for decelerating meteoroids is provided at \url{https://github.com/desertfireballnetwork/alpha_beta_modules}. 

$\alpha$ is related to the initial mass of the meteoroid ($M_0$, Equation \ref{eqn:me}) and the entry angle ($\gamma$), while $\beta$ is related to the instantaneous mass ($M_f$, Equation \ref{eqn:mf}) and the shape change coefficient ($\mu$) \citep{Lyytinen2016}:
\begin{align}
M_0 &= \frac{1}{2}\frac{c_d \rho_0 h_0 S_0}{\alpha \sin\gamma} = \left( \frac{1}{2}\frac{c_d A_0 \rho_0 h_0}{\alpha \rho_m^{2/3} \sin\gamma} \right)^3 \label{eqn:me}\\
M&=M_0 \, \exp\left \{-\cfrac{\beta}{1-\mu}\left(1-\left(\cfrac{V}{V_0}\right)^2\right) \right \}.\label{eqn:mf}
\end{align}
If quantitative values of these masses are required then assumptions must be made for the drag coefficient ($c_d$), initial cross sectional area ($S_0$) or initial shape coefficient ($A_0$) and meteoroid bulk density ($\rho_m$); the atmospheric surface density ($\rho_0$) is typically set to 1.21 kg/m$^3$. Applying such assumptions is similar to other methods, albeit the parameters that are needed to assume in this case have a limited range of values (meteoroid densities are well documented, as are shape, shape change and drag coefficients). $\beta$ here entirely replaces the need to assume an ablation parameter and subsequently a luminous efficiency -- the two most highly uncertain parameters usually required.
The advantage of this method, however, lies not in extracting individual parameters, but in assessing the relationship between $\alpha$ and $\beta$ values directly. 
With such a large data set, we wish to determine if any deductions can be made from groupings in these parameter spaces. 
By rearranging Equation \ref{eqn:me} for $\alpha$, we can see that a body of different entry masses, slopes and volumes are able to produce the same $\alpha$ values. 
The inclusiveness of these two parameters makes them more appropriate than the typical suite of parameters for predicting the outcomes of meteoroid atmospheric entry. 

\begin{figure}[h!]
    \centering
    \includegraphics[width=0.8\textwidth]{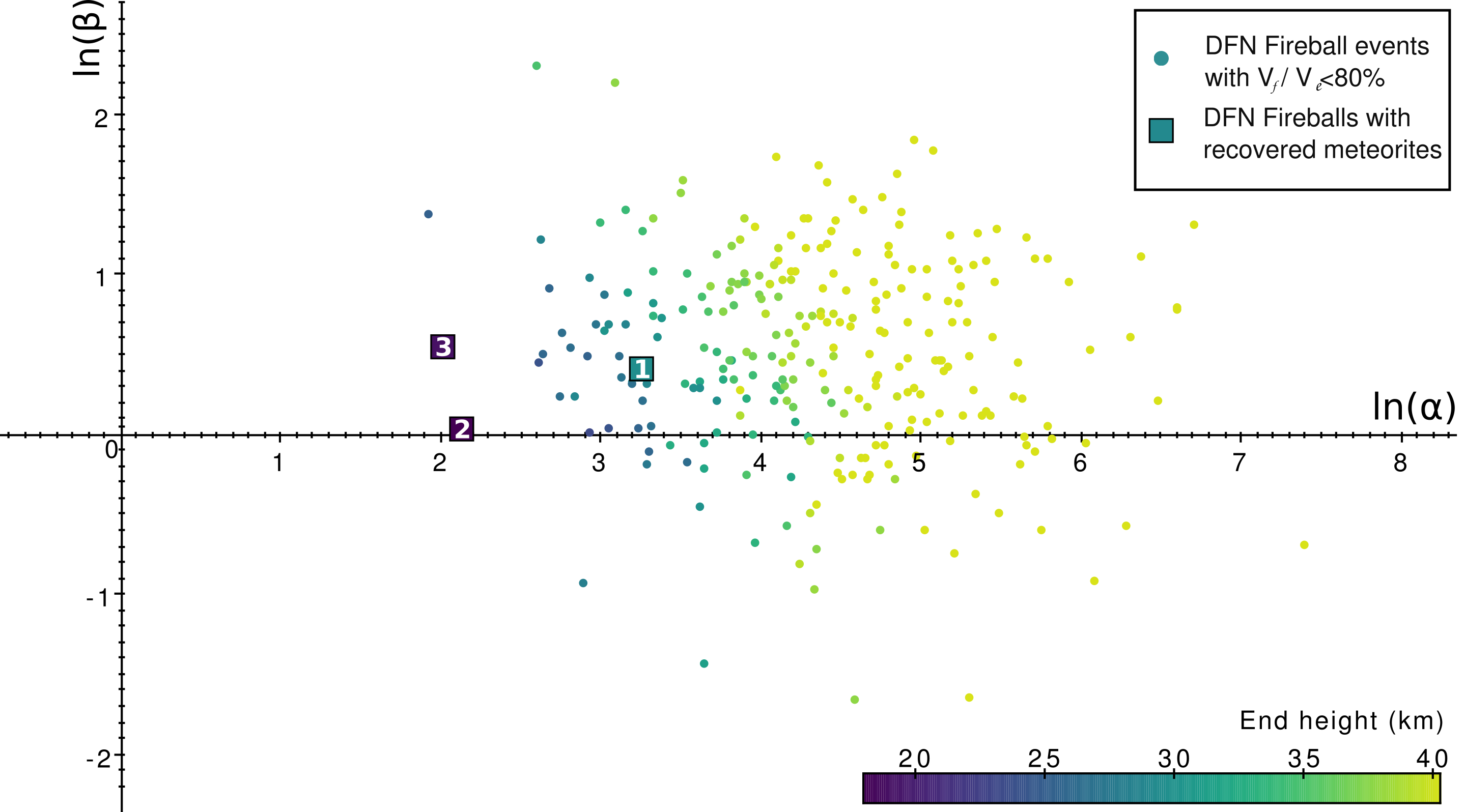}
    \caption{Distribution of $\alpha$ and $\beta$ parameters for Desert Fireball Network fireballs. Recovered meteorite falls plotted: (1) Bunburra Rockhole (DN200707B); (2) Murrili (DN151127\_01); Dingle Dell (DN 161031\_01)}
    \label{fig:DFN_data}
\end{figure}

We extracted all fireballs within the current DFN dataset where there is noticeable deceleration ($V_f/V_0<80\%$), and have calculated $\alpha$ and $\beta$ value for the resulting 278 events (see supplementary material for reduced data). 
We plot the results in a similar fashion to \citet{Gritsevich2012}, taking the natural logarithm of the $\alpha$ and $\beta$ values (Figure \ref{fig:DFN_data}). Although not a direct input parameter of either of Equation \ref{eqn:me}-\ref{eqn:mf}, the final observed height of the fireball (where the observation limit of the hardware can no longer observe ablation)
shows a clear horizontal trend with little relationship to $\beta$. Points with lower $\ln\alpha$ values will also have higher initial masses, as given by Equation \ref{eqn:me}.

\section{Determining the meteorite fall region}\label{sec:fall_region}
As previously stated, if we were to assume values for, say, density and shape in Equation \ref{eqn:me}, it is possible to then calculate the entry mass of a meteoroid using $\alpha$. Further assuming the shape change coefficient of the body can give a final mass using the $\beta$ value and Equation \ref{eqn:mf} (with luminosity values, $\mu$ can be determined following \citealp{bouquet2014simulation}). Here we plot a series of bounding curves for a given set of assumptions on the $\alpha$--$\beta$ diagram. This is an ideal visual tool for quickly assessing which fireballs from a large network might be meteorite droppers. 

\begin{figure}[h!]
    \centering
    \includegraphics[width=1\textwidth]{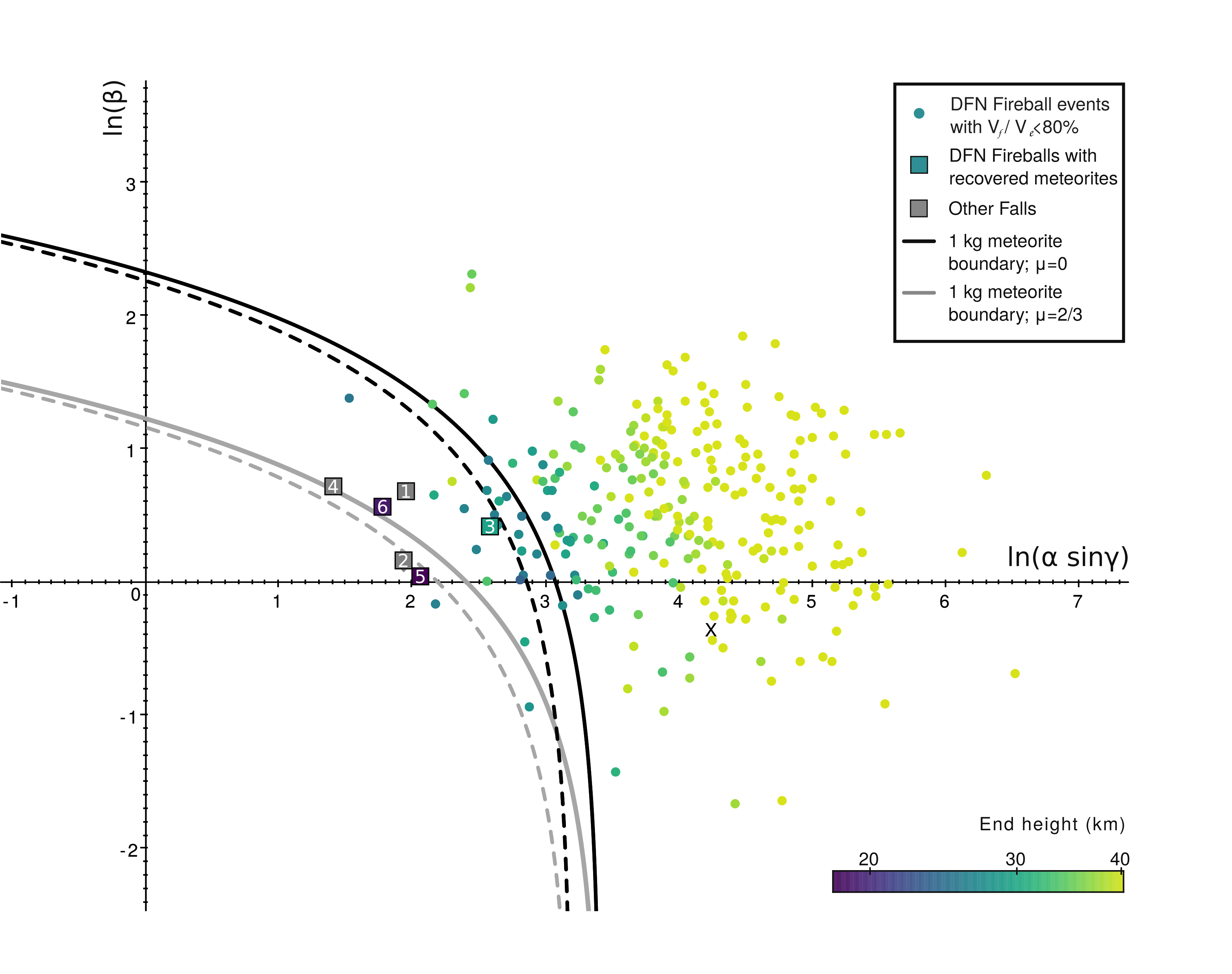}
    \caption{Distribution of fireballs from the Desert Fireball Network (DFN) with trajectory slope dependence removed (x-axis is now a function of $\gamma$). This changes the relationship between $\alpha$ and end height seen in Figure \ref{fig:DFN_data}. The bounding line for a 1 kg meteorite is shown in black for the case where there is no spin ($\mu=0$) and in grey where spin allows uniform ablation over the entire surface ($\mu=2/3$). Solid lines are for likely values of $c_d\,A=1.5$ and are dashed if $c_d\,A=1.21$. Meteorite falls plotted: (1) Innisfree (MORP285, 2.07 kg+); (2) Lost City (PN40590, 9.83 kg+); (3) Bunburra Rockhole (DN200707B, 174 g+); (4) Annama (FFN, 120 g); (5) Murrili (DN151127\_01, 1.68 kg); (6) Dingle Dell (DN161031\_01, 1.15 kg), where masses are given for largest recovered fragment and `+' indicates other fragments were found. Also note that the $\alpha$ -- $\beta$ values for Annama (4) were calculated using the method of \citet{Lyytinen2016} where a realistic atmosphere model is used rather than the exponential atmosphere as for other falls. 
    }
    \label{fig:main}
\end{figure}

As discussed in \citet{Gritsevich2012} the interpretation of the events is biased to the trajectory slope, individual for each event. Here we look at removing the effect of trajectory slope from the $\alpha$--$\beta$ diagram. If we plot instead $\ln(\alpha \,sin\gamma)$ as the x-axis, this effect is removed (Figure \ref{fig:main}). The clear horizontal trend in end heights, discussed in the previous section, now falls apart; there is no longer a distinct relationship. This is where the modified $\alpha$--$\beta$ diagram in Figure \ref{fig:main} is a more inclusive classification tool for fireballs. We no longer need to rely on final velocity and final end height requirements 
to classify a meteorite dropping event.

How are we then able to identify such a meteorite dropping region in these plots?
If we would like to assess the relationship between $\alpha$, $\beta$ and mass, we can extract $\alpha$ from Equation \ref{eqn:me} to give a parameter $M_0^*$ which is no longer dependent on $\alpha$ or the slope of the trajectory (Equation \ref{eqn:me_star}):
\begin{equation}
M_0 = \frac{1}{\alpha^3 \sin^3\gamma} M_0^* \, ,\qquad \mathrm{where} \qquad M_0^* = \Bigg( \frac{1}{2}\frac{c_d \rho_0 h_0 A_0}{\rho_m^{2/3} } \Bigg)^3.\label{eqn:me_star}
\end{equation}
To assess the final mass of a fireball, we look at Equation \ref{eqn:mf} in the case where the velocity becomes insignificant compared to the entry velocity (where $(V/V_0)^2 \longrightarrow 0$):
\begin{equation}
M_f=\frac{1}{\alpha^3 \sin^3\gamma} M_0^* \exp\left \{-\cfrac{\beta}{1-\mu} \right \}\label{eqn:mf_end}
\end{equation}
To define a region on the modified $\alpha$--$\beta$ diagram where a certain minimum final mass is obtainable, we can rearrange Equation \ref{eqn:mf_end} for $\beta$:
\begin{align}
\beta &= \left(\mu-1\right)\left(ln\left(\frac{M_f}{M_0^*}\right)+3 \ln\left(\alpha\sin\gamma\right)\right).\label{eqn:mu_}
\end{align}
To solve Equation \ref{eqn:mu_} 
for a final mass of $M_f=1$ kg, we use a density, $\rho_m=3500\,kg/m^3$ and a typical shape-drag coefficient, $c_dA=1.5$ \citep{Gritsevich2008Validity} to get a value of $ln(M_f/M_0^*) = -10.21$. 
We can plot this boundary line given the two extreme values of the shape change coefficient -- when $\mu = 0$, there is no spin of the meteoroid, and when $\mu = 2/3$, there is sufficient spin to allow equal ablation over the entire meteoroid surface and no shape change is expected to occur, giving:
\begin{align}
\mu = 0\,, \qquad \ln\beta &= \ln\{10.21-3\, \ln\left(\alpha\sin\gamma\right)\}\label{eqn:mu0_br}\\
\mu = \frac{2}{3}\,, \qquad \ln \beta &= \ln\{3.4- \ln\left(\alpha\sin\gamma\right)\}.\label{eqn:mu23_br}
\end{align}
These boundary curves are plotted on the modified $\alpha$ -- $\beta$ diagram in Figure \ref{fig:main} for such a 1 kg mass. 
Many similar scenarios can be actualised for various shapes, densities and minimum terminal mass values\footnote{The interactive tool available at \url{https://github.com/desertfireballnetwork/alpha_beta_modules} provides a means to investigate these scenarios}. Such an example plotted in Figure \ref{fig:main} includes using $c_dA=1.21$ for a perfectly spherical meteoroid body. 

As mentioned previously, there is a general rule of thumb that crudely uses a fireball end height of $<35$ km and terminal velocity $<10$ km s$^{-1}$ to determine which meteoroids may have survived entry. 
If we define a macroscopic meteorite-dropping event as having a final mass of $>50$ g (following \citealp{Halliday1996} and \citealp{gritsevich2011DokPh..56..199G}), Equations \ref{eqn:mu0_br}-\ref{eqn:mu23_br} become:
\begin{align}
\mu = 0\,, \qquad \ln\beta &= \ln\{13.20-3\, \ln\left(\alpha\sin\gamma\right)\}\label{eqn:main_a}\\
\mu = \frac{2}{3}\,, \qquad \ln \beta &= \ln\{4.4- \ln\left(\alpha\sin\gamma\right)\},\label{eqn:main_b}
\end{align}
given a $\rho_m=3500\,kg/m^3$ and a $c_dA=1.5$.

\begin{figure}[h!]
    \centering
    \includegraphics[width=1\textwidth]{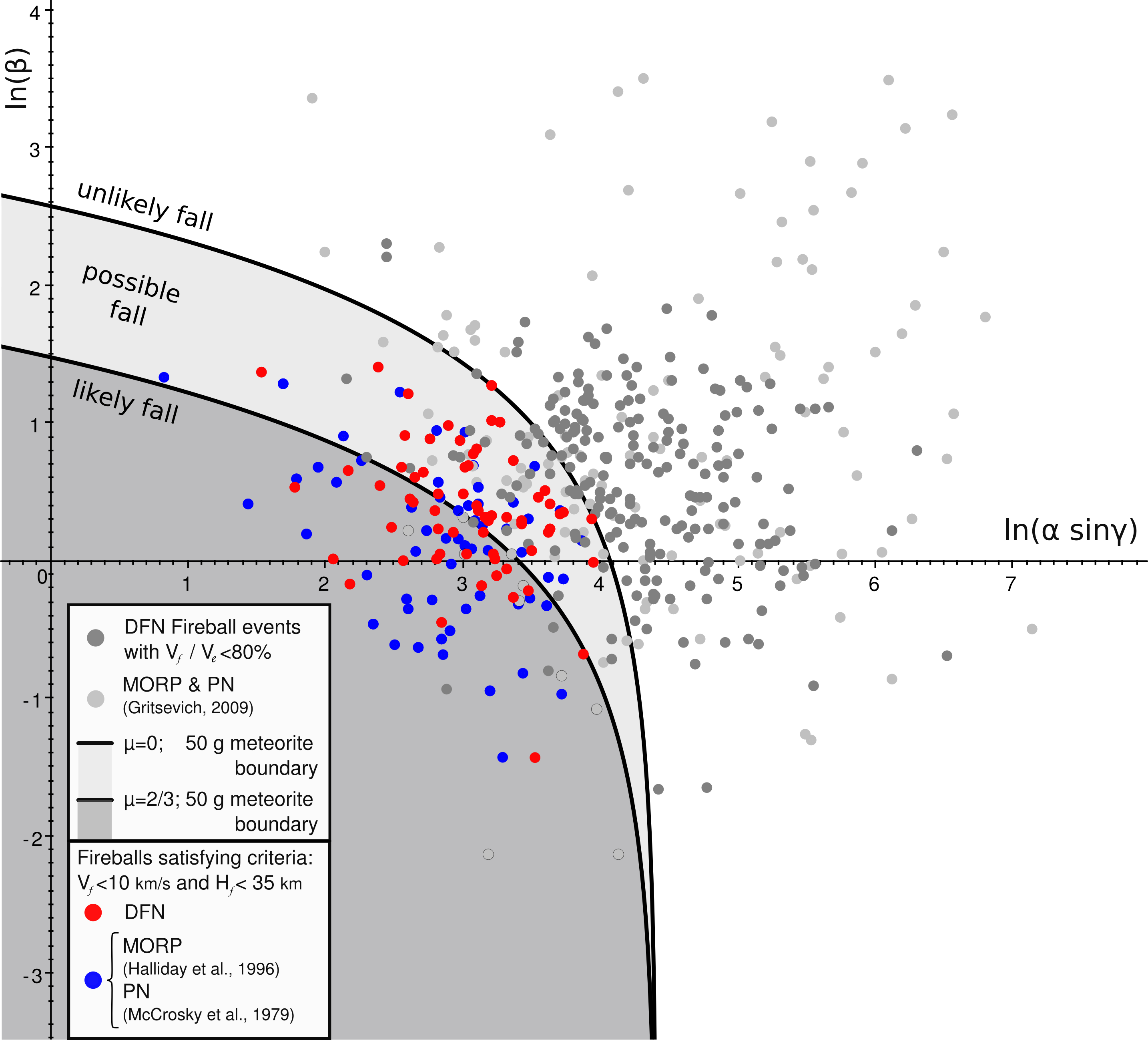}
    \caption{Distribution of fireballs from both the Desert Fireball Network (DFN) and previous studies (Meteor Observation and Recovery Project \citealp{Halliday1996}; Prarie Network \citealp{mccrosky1979prairie}). Fireball events that meet the criteria $V_f<10$ km s$^{-1}$ and $H_f<35$ km are considered likely meteorite droppers (after \citealp{brown2013meteorites}) and are shown in red (DFN) and blue (previous studies). Boundary lines for a 50 g meteorite are given for the two extremes of the shape change coefficient $\mu$ using Equations \ref{eqn:main_a}-\ref{eqn:main_b}. The area beyond both these lines will be unlikely to drop a $>50$ g meteorite, while those within the dark grey `likely fall' region will be strong meteorite-producing candidates.}
    \label{fig:crit}
\end{figure}

In Figure \ref{fig:crit} we plot these boundary curves with the fireball data from the DFN and these previous studies (MORP \& PN).
Note that PN and MORP data were not subject to the same deceleration thresholding applied to DFN data here, and any differences in $\alpha$ -- $\beta$ values for these other studies to \citet{Gritsevich2012} are due to the slope dependence being addressed here. 
As the boundary lines are given for the two extremes of the shape change coefficient $\mu$, events falling beyond the $\mu=0$ line are unlikely to have produced a 50 g meteorite. Fireballs associated with known meteor shower events all plot in this area, with high $\ln(\beta)$ and $\ln(\alpha\, \sin(\gamma))$ values. Fireballs below the $\mu=2/3$ line are strong meteorite-producing candidates. 
The significant area between these two curves illustrates the sensitivity of the dynamic flight equations to meteoroid rotation. As a subsequent step, the shape change coefficient can be calculated for individual events from luminosity values following \citet{bouquet2014simulation}.

Events that meet the empirical criteria ($V_f<10$ km s$^{-1}$ and $H_f<35$ km) are highlighted in Figure \ref{fig:crit}.
Within the `likely fall' area, nearly all events meet this criteria. All highlighted events fall within the $\mu=0$ bounding line. 
These bounding lines are highly compatible with the empirical fall criteria and present a physical 
basis for the classification of such events.
We propose that these bounding lines be used in future for more rigorously determining a meteoroid's potential to survive entry. 
We will further discuss the advantages and limitations of using the $\alpha$ -- $\beta$ diagram, and the cases in particular of `likely fall' events that do not meet the empirical criteria. 

\section{Discussion}\label{sec:discussion}
Figure \ref{fig:crit} clearly demonstrates the suitability of Equations \ref{eqn:main_a}-\ref{eqn:main_b} to determine the likelihood of a macroscopic terminal mass. 
Although the general rule of thumb 
is consistent, there are multiple events in both the 
`possible fall' region
and the `likely fall' region that do not satisfy the simplified empirical criteria. 
Could these missed events really be falls? Let us first discuss the possible limitations of this method before addressing these events. 

Once an event is located on this modified $\alpha-\beta$ diagram, if it falls in either of the grey regions in Figure \ref{fig:crit} it is worth further investigation. 
Following this $\alpha$ -- $\beta$ approach, there are several advancements on this basic implementation that can be performed. 
Despite using the simplified exponential atmosphere as a generic model, the actual atmospheric conditions for individual cases can be accounted for, given the time and location of the fireball as described in \citet{Lyytinen2016}. 
There is also a strong sensitivity of this method to the initial velocity, as the normalisation of velocity values uses $V_0$. Although a first order $V_0$ can be used initially, for possible fall events, it is best to recalculate velocities using a robust method (such as discussed in \citealp{Sansom2015} and \citealp{vida2018modelling}).
Differences in $V_0$ calculation methods by MORP and PN could be a possible explanation for many of the light grey events falling in the `likely fall' region. 
Using more realistic atmospheric conditions \citep{Lyytinen2016}, and with recalculated $V_0$ values, the resulting $\alpha$ and $\beta$ values become more representative.

The position of an event on the $\alpha$ -- $\beta$ diagram within the grey region indicates that there may be a macroscopic mass at the last observation point. This may not, in some cases, correspond to the terminal bright flight mass, or to an equivalent meteorite mass on the ground. For example when the last observed point is not the end of the bright flight trajectory, due to missing observations, or distance of the trajectory end to the observer. Distant fireballs may continue to ablate beyond the limiting magnitude of imaging systems. MORP and PN studies used large format film systems recording a single image per night, with fireball segments recorded at a frequency of 4 Hz \cite{halliday1978} and 20 Hz \cite{McCrosky1965Prairie} respectively. PN systems identify typical projected limiting magnitudes of -3 at the centre of their frames (with -5 toward the edges) \cite{McCrosky1965Prairie}. These systems may not have been sensitive enough to reliably image the end of bright flight. Such missing information could account for why terminal masses may appear overestimated in the $\alpha$ -- $\beta$ diagram.
Fragmentation within the bright flight is to some extent accounted for by the nature of fitting the deceleration profile with Equation \ref{eqn:y}. Where fragmentation occurs at the end of the bright flight however, the terminal mass expected will no longer be a single main mass. Modelling of fragments through darkflight may still be valuable if the end mass is significant enough. 
An estimate of this terminal mass can be calculated using Equation \ref{eqn:mf}.  
This does require assumptions to be made for density, shape and of course $\mu$. For a more in depth analysis/assessment of specific meteoroid trajectories, more involved modelling techniques such as \citet{sansom20193d} and \citet{egal2017challenge} can now be applied with confident use of resources. 

Let us return to the grey DFN events in Figure \ref{fig:crit} that are within the `likely fall' region (we include the two on the $\mu=2/3$ line). Of the five, the most eye catching is at [2.88,-0.936] in Figure \ref{fig:crit} and from video data shows significant flaring, including a final late flare. The mass at this point is still significant (~1 kg) and a search for fragments will be conducted in the future. The event at [2.30, 0.75] in Figure \ref{fig:crit} is a great example of hardware limitations interfering with expected results. DFN observatories are designed to take a 25 second long-exposure image every 30 seconds. This 5 second down time allows images to be saved and systems to be reset. This event likely continued to ablate beyond the end of the exposure and was unfortunately not captured in the subsequent image. 
The remaining three are triangulated from observatories at significant ranges; the closest camera to DN151105\_15 (Figure \ref{fig:crit} [3.08,0.27]) was 430 km. 
These are therefore still possible fall candidates that were missed by the empirical criteria, simply because the end of bright flight was not observed. These were modelled using \citet{Sansom2015} and masses at this last observed point are all $>100$ g.
This method is therefore able to identify likely fall events that might previously have been missed if using the empirical criteria for a typical meteorite-dropping event. 

\section{Conclusions}\label{sec:conc}
Here we demonstrate an $\alpha$ -- $\beta$ diagram as a simple, yet powerful, tool to visualise which fireball events are likely to have macroscopic terminal masses. 
We plot 278 fireballs from the Desert Fireball Network on a modified $\alpha$ -- $\beta$ diagram, accounting for the differences in trajectory slopes (Figure \ref{fig:main}). 
Boundary lines can be plotted to define a region of events having a given minimum terminal mass. The shape change coefficient, $\mu$, is capable of enhancing mass loss and its influence should be considered.
Equations \ref{eqn:main_a}-\ref{eqn:main_b} define the boundary curves for a terminal 50 g chondritic mass, given the two extremes of meteoroid rotation ($0<\mu<2/3$; Figure \ref{fig:crit}). Events beyond both these lines are unlikely to have survived atmospheric entry, while those below both lines are likely to have dropped a macroscopic meteorite. Depending on the meteoroid rotation, events in the region between these lines should also be considered as possible falls. 
Events from previous studies (MORP and PN) 
are also shown for comparison.

Events that meet the current empirical fall criteria ($V_f<10$ km s$^{-1}$ and $H_f<35$ km) all lie within the proposed fall regions of the $\alpha$ -- $\beta$ diagram (Figure \ref{fig:crit}). Not only can this method locate all events identified by the empirical criteria, but it is able to provide the physical justification for highlighting such events. 
Additionally, the $\alpha$ -- $\beta$ method is able to detect likely fall events that do not meet these empirical criteria, identifying non-typical events. The use of the $\alpha$ -- $\beta$ criterion is a  way to quickly and easily identify key events in large datasets. This method is easily automated and has previously been shown to scale to airburst and cratering events. With more data, this could become increasingly useful for identifying where hazardous material may be originating from in the solar system.

\section{Acknowledgements and author notes}
EKS acknowledges the Australian Research Council for funding received as part of the Australian Discovery Project scheme (DP170102529). 

SSTC authors acknowledge  institutional support from Curtin University.

MG acknowledges Academy of Finland project no. 325806 and the Russian Foundation for Basic Research, project nos. 18-08-00074 and 19-05-00028. Research at the Ural Federal University is supported by the Act 211 of the Government of the Russian Federation, agreement No 02.A03.21.0006.

This research made use of TOPCAT for visualisation and figures \citep{taylor2005topcat}.

The code used to determine the $\alpha$ and $\beta$ parameters for a fireball data set (after \citealp{Gritsevich2009parameters}) is available on GitHub as an interactive Jupyter notebook (\url{https://github.com/desertfireballnetwork/alpha\_beta\_modules}).

\section{Summary of definitions and abbreviations}\label{sec:defs}
\begin{longtable}{p{1cm} c p{12cm}}
$A_0$	&	 $-$ 	&	Initial shape factor - a cross sectional area to volume ratio $A=S\left(\frac{\rho_m}{m}\right)^{2/3}$.	\\
$c_d $ 	&	 $-$ 	&	 Drag coefficient.	\\
$c_h $ 	&	 $-$ 	&	 Heat-transfer coefficient.	\\
$\Bar{Ei}$      &	 $-$ 	&	 Exponential integral, $\Bar{Ei}(x)=\int_{\infty}^{x}\frac{e^{z}}{z}\,dz\,$. \\
$\mathbf{g}$ 	&	 $-$ 	&	 Vector of local gravitational acceleration ($m\, s^{-2}$).	\\
$h_0$ 	&	 $-$ 	&	 Scale height of the homogeneous atmosphere ($h_0=7160\,m$).	\\
$H^*$ 	&	 $-$ 	&	 Enthalpy of sublimation ($J\, kg^{-1}$).	\\
$m$ 	&	 $-$ 	&	 Normalised meteoroid mass, $m = \cfrac{M}{M_0}$  (dimensionless).	\\
$M$ 	&	 $-$ 	&	 Meteoroid mass ($kg$).	\\
$M_0$ 	&	 $-$ 	&	 Initial entry mass of meteoroid at the beginning of the observed, luminous trajectory ($kg$).	\\
$M_0$ 	&	 $-$ 	&	 An intermediate variable defined by Equation \ref{eqn:me_star} (dimensionless).\\
$M_f$ 	&	 $-$ 	&	 Terminal mass of the main meteoroid body at the end of the luminous trajectory ($kg$).	\\
$S$ 	&	 $-$ 	&	 Cross sectional area of the body ($m^2$).	\\
$S_0$ 	&	 $-$ 	&	 Initial cross sectional area of the body ($m^2$).	\\
$v$ 	&	 $-$ 	&	 Normalised meteoroid velocity, $v = \cfrac{V}{V_0}$  (dimensionless).	\\
$V$ 	&	 $-$ 	&	 Meteoroid velocity ($m\, s^{-1}$).	\\
$V_0$ 	&	 $-$ 	&	 Initial entry velocity of the meteoroid at the beginning of the observed, luminous trajectory ($m\, s^{-1}$).	\\
$V_f$ 	&	 $-$ 	&	 Terminal velocity of the main meteoroid body at the end of the luminous trajectory ($m\, s^{-1}$).	\\
$y$ 	&	 $-$ 	&	 Normalised meteoroid height, $y = \cfrac{altitude}{h_0}$  (dimensionless).	\\
$\alpha$	&	 $-$ 	&	Ballistic Coefficient.	\\
$\beta$	&	 $-$ 	&	Mass loss parameter.	\\
$\gamma $ 	&	 $-$ 	&	Angle of the meteoroid flight to the horizontal.	\\
$\mu$	&	 $-$ 	&	Shape change coefficient representing the rotation of a meteoroid body ($0<\mu<2/3$).	\\
$\rho_a$ 	&	 $-$ 	&	 Atmospheric density ($kg\, m^{-3}$).	\\
$\rho_m$ 	&	 $-$ 	&	Meteoroid bulk density ($kg\,m^{-3}$).	\\
\end{longtable}

\section{References}
    \bibliography{Bibliography}{}
    \bibliographystyle{abbrvnat}

\end{document}